\documentclass[aps,prd]{revtex4}
\usepackage{graphicx}
\usepackage{latexsym}
\usepackage{amsmath}
\usepackage{amsfonts}
\usepackage{amssymb}
\usepackage{verbatim}

\newcommand{\be}{\begin{equation}}
\newcommand{\ee}{\end{equation}}
\newcommand{\bea}{\begin{eqnarray}}
\newcommand{\eea}{\end{eqnarray}}

\newcommand{\pa}{\partial}

\newcommand{\bb}{\bibitem}
\def\pls{\partial\!\!\!/}
\def\bb{\bibitem}
\def\as{A\!\!\!/}

\def\ps{p\!\!\!/}
\def\bs{b\!\!\!/}

\def\ds{\partial\!\!\!/}

\def\bb{\bibitem}
\newcommand{\ben}{\begin{eqnarray}}
\newcommand{\een}{\end{eqnarray}}

\begin{document}

\title{On the effective action of the vacuum photon splitting in Lorentz-violating QED}
\author{F. A. Brito, E. Passos and P. V. Santos} 
\affiliation{Departamento de F\'\i sica,
Universidade Federal de Campina Grande, Caixa Postal 10071,
58109-970  Campina Grande, Para\'\i ba,
Brazil
}


\begin{abstract}
We consider one-loop radiative corrections from Lorentz- and CPT- violating extended QED to  address the specific problem of finding explicitly an effective action describing amplitude of photon triple splitting. We show that it is not possible to find a nonzero photon triple splitting effective action, at least by using the derivative expansion method (at zero external momenta), up to leading order in the Lorentz- and CPT-  violating parameter.
\end{abstract}
\pacs{XX.XX, YY.YY} \maketitle


\section{Introduction}

The main purpose of the present work is to discuss the one-loop radiative corrections from Lorentz- and CPT-violating extended QED to
investigate the induction of a nonzero amplitude for vacuum photon splitting \cite{kp2003}  by using the derivative expansion 
of fermion determinants. We will address the specific problem of finding explicitly an effective action describing amplitude of  photon triple splitting. In this case, we search for the possibility of generating
quantum radiatively a nonzero term responsible for this splitting  even though it is not present in the bare Lagrangian. We will follow the same lines of inducing Chern-Simons-like actions --- strongly constrained through  several measurements such as astrophysical birefringence, Schumann resonances and CMB polarization \cite{cfj,kost2008}. 

The induction of a photon triple splitting term is
an important result in the study of the Lorentz symmetry violation initiated in \cite{kp2003}, though an effective action was not found there. As far as we know this is still an open problem that justifies an investigation of this issue by some well-known method. We shall consider the derivative expansion method in the present study.
The aforementioned  term may naturally emerge as a perturbative correction in Lorentz- and CPT-violating extended QED theory suggested in \cite{Colladay:1998fq}. Let us now focus only on one part of such extension by considering 
the QED extended with an axial vector term
\bea\label{01}
{\cal L}=\bar\psi(i\pls-M_{b})\psi-e\bar\psi\as\psi,
\eea
where $M_{b}\equiv m +\bs\gamma_{5}$ with $m$ being the single mass of the fermion field and $b_{\mu}$  a 
constant four-vector that controls the Lorentz- and CPT violation in the fermion sector. To calculate
the CPT-odd effective action for photon triple splitting, we shall integrate out the massive fermion field in the functional integral as discussed below.


\section{The effective action for photon triple splitting}
We can integrate out the fermion field $\psi$ in the
functional integral to obtain the effective action for the gauge field $A_{\mu}(x)$ that reads \cite{3}
\bea\label{02}
S_{eff}[b,A]= -i{\rm\, Tr\, ln}\big[\ps- M_{b}-e\as\big].
\eea
Here the symbol Tr stands for the trace over Dirac matrices, trace over the internal space
as well as for the integrations in momentum and coordinate spaces.
The first nondynamical determinant factor has been absorbed into normalization of
the path integral such that
\bea\label{03}
S_{eff}[b,A]=S_{eff}[b]+S_{eff}^{(n)}[b,A],
\eea
with
\bea\label{04}
S_{eff}^{(n)}[b,A]&=&i{\rm Tr}\ln\Big[1-\frac{e}{\ps-M_{b}}\as(x)\Big],\nonumber\\&=&
i{\rm Tr}\sum^{\infty}_{n=1}\frac{1}{n}\Big[-i e\, G^{b}_{F}(p)\as(x) \Big]^{n}.
\eea
Let us now consider the first four terms of the power expanded logarithm in Eq.(\ref{04}) to write
\bea\label{06}
S_{eff}^{(n)}[b,A]&=&i{\rm Tr}\Big[-ie\, G^{b}_{F}(p)\as(x)-\frac{e^{2}}{2}G^{b}_{F}(p)\as(x) G^{b}_{F}(p)\as(x)+\frac{ie^{3}}{3}G^{b}_{F}(p)\as(x)G^{b}_{F}(p)\as(x)G^{b}_{F}(p)\as(x)
\nonumber\\&-&\frac{e^{4}}{4}G^{b}_{F}(p)\as(x)G^{b}_{F}(p)\as(x)G^{b}_{F}(p)\as(x)G^{b}_{F}(p)\as(x)+\cdot\cdot\cdot\Big],
\eea
where $G^{b}_{F}(p)$ is the exact fermion propagator expressed in the form
\bea\label{05}
G_{F}^{b}(p)=\frac{i}{\ps-M_{b}}.
\eea
In the expansion (\ref{06}) one can recognize the first term as a tadpole, the second is known to induce the Chern-Simons-like term, the third term does not contribute and finally the fourth term is the one that  will search for an induced photon triple splitting term in the effective action. All of these terms are one-loop contributions. In the following we shall focus only on the fourth term.

The effective action for photon triple splitting can be obtained from the following contribution
\bea\label{07}
S_{eff}^{(4)}[b,A]&=&-\frac{ie^{4}}{4}{\rm Tr}\Big[G_{F}^{b}(p)\as(x)G_{F}^{b}(p)\as(x)G_{F}^{b}(p)\as(x)G_{F}^{b}(p)\as(x)\Big].
\eea
Now applying the main property of derivative expansion method, we observe that any function of momentum can be converted into a coordinate dependent quantity as \cite{ed}
\bea\label{08}
A(x)f(p)=(f(p-i\pa)A(x)).
\eea
The parenthesis on the right hand side merely emphasizes that the derivatives act
only on $A(x)$. In this case, 
the photon triple splitting contribution action will become
\bea\label{09}
S_{eff}^{(4)}[b,A]&=&-\frac{ie^{4}}{4}{\rm Tr}\Big[G_{F}^{b}(p)\gamma^{\mu}(G_{F}^{b}(p-i\pa)A_{\mu})\gamma^{\nu}(G_{F}^{b}(p-i\pa)A_{\nu})\gamma^{\beta}(G_{F}^{b}(p-i\pa)A_{\beta})\as\Big].
\eea
By using the following expansion
\bea\label{10}
G_{F}^{b}(p-i\pa)A_{\mu}&=&\frac{i}{\ps-i\ds-M_{b}}A_{\mu}\nonumber\\&=&G_{F}^{b}(p)A_{\mu}+G_{F}^{b}(p)\gamma^{\rho} G_{F}^{b}(p)(\pa_{\rho}A_{\mu})+\cdot\cdot\cdot,
\eea
up to first order in the derivative and the cyclic properties of the trace of the
product of matrices, we rewrite the expression (\ref{09}) in the form
\bea\label{efa1}
&&S_{eff}^{(4)}[b,A]=-\frac{ie^{4}}{4}\int d^{4}x (\pa_{\rho}A_{\mu})\int \frac{d^{4}p}{(2\pi)^{4}}\Big[3{\rm tr}\big[G_{F}^{b}(p)\gamma^{\rho}G_{F}^{b}(p)\as G_{F}^{b}(p)\as G_{F}^{b}(p)\as G_{F}^{b}(p)\gamma^{\mu}\big]+\nonumber\\&&2{\rm tr}\big[G_{F}^{b}(p)\gamma^{\rho}G_{F}^{b}(p)\as G_{F}^{b}(p)\as G_{F}^{b}(p)\gamma^{\mu}G_{F}^{b}(p)\as\big]+{\rm tr}\big[G_{F}^{b}(p)\gamma^{\rho}G_{F}^{b}(p)\as G_{F}^{b}(p)\gamma^{\mu} G_{F}^{b}(p)\as G_{F}^{b}(p)\as\big]\Big],
\eea
where the symbol ${\rm tr}$ denotes the trace of the product of the
gamma matrices. 
\section{Massive Fermion Propagator}
In the massive theory, the exact fermion propagator may be read off directly
from the bare Lagrangian density and is given in  (\ref{05}). We may rationalize such propagator to obtain \cite{PV,Chung,AA}
\bea\label{pr}
G_{F}^{b}(p)=i\frac{\big(\ps+m-\bs\gamma_{5}\big)\big(p^{2}-m^{2}-b^{2}+\big[\ps,\bs\big]\gamma_{5}\big)}{\big(p^{2}-m^{2}-b^{2}\big)^{2}+4\big[p^{2}b^{2}-(b\cdot p)^{2}\big]}.
\eea
This form of the propagator represents one of the
ways in which the presence of $b$ will affect the photon splitting effect.
However, the direct application of this exact form into the expression (\ref{efa1}) is quite involved, so that we 
expand the expression (\ref{efa1}) as a power series in $b$, an
approximation developed in \cite{3}, i.e.,  we expand the
exact fermion propagator $G_{F}^{b}(p)$ and take the leading order in $b$ as in the form
\bea\label{09.2}
G_{F}^{b}(p)=G_{F}(p)+G_{F}(p)(-i\bs\gamma_{5})G_{F}(p)+\cdot\cdot\cdot\;\;\;{\rm with}\,\,\;\;G_{F}(p)=\frac{i}{\ps-m}.
\eea
Let us now use the above expansion into expression (\ref{efa1}) and thereafter, we can
compute the trace of gamma matrices to isolate the effective action as follows
\bea\label{10.a}
S_{eff}^{(4)}[b,A]=2\int d^{4}x\big[\Pi_{a}^{\mu\rho\nu\alpha\beta}+\Pi_{b}^{\mu\rho\nu\alpha\beta}+\Pi_{c}^{\mu\rho\nu\alpha\beta}+
\Pi_{d}^{\mu\rho\nu\alpha\beta}\big](\pa_{\rho}A_{\mu})A_{\nu}A_{\alpha}A_{\beta},
\eea
where $\Pi_{a}^{\mu\lambda\nu\alpha\beta}$, $\Pi_{b}^{\mu\lambda\nu\alpha\beta}$, $\Pi_{c}^{\mu\lambda\nu\alpha\beta}$ and $\Pi_{d}^{\mu\lambda\nu\alpha\beta}$ are self-energy tensors written as below
\bea\label{15}
&&\Pi_{a}^{\mu\rho\nu\alpha\beta}=5ie^{4}b_{\lambda}\epsilon^{\mu\rho\nu\lambda}\int\frac{d^{4}p}{(2\pi)^{4}}\frac{(p^{2}-m^{2})\eta^{\alpha\beta}-6p^{\alpha}p^{\beta}}{(p^{2}-m^{2})^{4}},
\nonumber\\&&
\Pi_{b}^{\mu\rho\nu\alpha\beta}=6ie^{4}b_{\lambda}\epsilon^{\mu\sigma\rho\lambda}\int\frac{d^{4}p}{(2\pi)^{4}}\frac{3(p^{2}-m^{2})p_{\sigma}p^{\nu}\eta^{\alpha\beta}-8p_{\sigma}p^{\nu}p^{\alpha}p^{\beta}}{(p^{2}-m^{2})^{5}},
\nonumber\\&&
\Pi_{c}^{\mu\rho\nu\alpha\beta}=6ie^{4}b_{\lambda}\epsilon^{\sigma\nu\rho\lambda}\int\frac{d^{4}p}{(2\pi)^{4}}\frac{(p^{2}-m^{2})p_{\sigma}p^{\mu}\eta^{\alpha\beta}-8p_{\sigma}p^{\mu}p^{\alpha}p^{\beta}}{(p^{2}-m^{2})^{5}},
\nonumber\\&&
\Pi_{d}^{\mu\rho\nu\alpha\beta}=6ie^{4}b_{\lambda}\epsilon^{\mu\nu\sigma\lambda}\int\frac{d^{4}p}{(2\pi)^{4}}\frac{(p^{2}-m^{2})p_{\sigma}p^{\rho}\eta^{\alpha\beta}-8p_{\sigma}p^{\rho}p^{\alpha}p^{\beta}}{(p^{2}-m^{2})^{5}}.
\eea
We can see that the above integrals are convergent by momentum power counting.
Now we use the following 4D-momentum formulas in the Minkowski space
\bea\label{I.1}
\int\frac{d^{4}p}{(2\pi)^{4}}\frac{1}{(p^{2}-m^{2})^{n}}&=&\frac{(-)^{n}}{16\pi^{2}}\frac{\Gamma(n-2)}{\Gamma(n)}\frac{i}{(m^{2})^{n-2}}
\nonumber\\&\stackrel{n=3}{\rightarrow}&
\frac{-i}{16\pi^{2}}\frac{1}{2m^{2}},
\eea
\bea\label{I.2}
\int\frac{d^{4}p}{(2\pi)^{4}}\frac{p^{\lambda}p^{\nu}}{(p^{2}-m^{2})^{4}}&=&\frac{(-)^{n-1}}{16\pi^{2}}\frac{\eta^{\lambda\nu}}{2}\frac{\Gamma(n-3)}{\Gamma(n)}\frac{i}{(m^{2})^{n-3}}
\nonumber\\&\stackrel{n=4}{\rightarrow}&\frac{-i }{16\pi^{2}}\frac{\eta^{\lambda\nu}}{12m^{2}},
\eea
and
\bea\label{I.3}
\int\frac{d^{4}p}{(2\pi)^{4}}\frac{p^{\sigma}p^{\nu}p^{\alpha}p^{\beta}}{(p^{2}-m^{2})^{n}}&=&\frac{(-)^{n}}{16\pi^{2}}\frac{G^{\sigma\nu\alpha\beta}}{4}\frac{\Gamma(n-4)}{\Gamma(n)}\frac{i}{(m^{2})^{n-4}}\nonumber\\&\stackrel{n=5}{\rightarrow}&\frac{-i }{16\pi^{2}}\frac{G^{\sigma\nu\alpha\beta}}{96m^{2}},
\eea
where $G^{\sigma\nu\alpha\beta}=\eta^{\sigma\nu}\eta^{\alpha\beta}+\eta^{\sigma\alpha}\eta^{\nu\beta}+\eta^{\sigma\beta}\eta^{\nu\alpha}$. One can notice that, 
by using the equations (\ref{I.1}), (\ref{I.2}) and (\ref{I.3}) into equations (\ref{15}), we have that all the tensors $\Pi_{a}^{\mu\rho\nu\alpha\beta}$, $\Pi_{b}^{\mu\rho\nu\alpha\beta}$, $\Pi_{c}^{\mu\rho\nu\alpha\beta}$ and $\Pi_{d}^{\mu\rho\nu\alpha\beta}$ independently {\it vanish}. Thus we cannot isolate a nonzero  photon triple splitting effective action through derivative expansion up to leading order in $b$.
\section{Conclusions}
In summary in the Lorentz- and CPT-violating extended QED is {\it not} possible to find a nonzero photon triple splitting effective action, at least by using the derivative expansion method (at zero
external momenta) up to leading order in $b$. This seems to be in accord with the result put forward in Ref.~\cite{kp2003}, since that result depends explicitly on the external momenta and then disappears for zero external momenta. Considering higher orders in $b$ is out of the scope of this short note but it should be addressed elsewhere --- same can also be done for photon decay \cite{cosm1}.  The massless case $m=0$, can also be addressed elsewhere. We anticipate that in this case the exact fermion propagator is easily rationalized  \cite{PV, alt1, baeta},
which may simplify the algebra in the calculation of expression (\ref{efa1}).
This study may follow the lines of those in the Chern-Simons-like action of extended massless QED whose induced coefficient is well defined and finite \cite{n1,n2}.





{\acknowledgments} We would like to thank the organizers and participants of ``Workshop Teoria Qu\^antica de Campos, Gravita\c c\~ao e Aspectos Computacionais'', Bras\'\i lia - Brazil, November 2010, for discussions and CNPq, PNPD-CAPES, PROCAD-NF/2009-CAPES for partial financial support.



\begin{thebibliography}{99}

\bibitem{kp2003}
  V.~A.~Kostelecky and A.~G.~M.~Pickering,
  Phys.\ Rev.\ Lett.\  {\bf 91}, 031801 (2003)
  [arXiv:hep-ph/0212382].

\bb{cfj} S. Carroll, G. Field, and R. Jackiw, Phys. Rev. D {\bf41}, 1231 (1990).

\bibitem{kost2008}
  V.~A.~Kostelecky and N.~Russell,
  Rev.\ Mod.\ Phys.\  {\bf 83}, 11 (2011)
  [arXiv:0801.0287 [hep-ph]].

\bibitem{Colladay:1998fq}
  D.~Colladay and V.~A.~Kostelecky,
  Phys.\ Rev.\  D {\bf 58}, 116002 (1998)
  [arXiv:hep-ph/9809521].

\bb{3} R. Jackiw and V. A. Kosteleck\'y, Phys. Rev. Lett. {\bf 82}, 3572 (1999).

\bb{ed}I. J. R. Aitchison and C. M. Fraser, Phys. Lett. B {\bf146},
63 (1984); Phys. Rev. D {\bf31}, 2605 (1985); C. M. Fraser, Z. Phys.
C {\bf28}, 101 (1985); A. I. Vainshtein, V. I. Zakharov, V. A.
Novikov, and M. A. Shifman, Yad. Fiz. (Sov. J. Nucl. Phys.) {\bf
39}, 77 (1984); J. A. Zuk, Phys. Rev. D {\bf 32}, 2653 (1985); M. K.
Gaillard, Nucl. Phys. B {\bf 268}, 669 (1986); A. Das and A. Karev,
Phys. Rev. D {\bf 36}, 623 (1987); K. S. Babu, A. Das, and P.
Panigrahi, Phys. Rev. D {\bf 36}, 3725 (1987);
  J.~R.~Nascimento, E.~Passos, A.~Y.~Petrov and F.~A.~Brito,
  JHEP {\bf 0706}, 016 (2007)
  [arXiv:0705.1338 [hep-th]].


\bb{PV} M. Perez-Victoria, Phys. Rev. Lett. {\bf 83}, 2518 (1999).

\bb{Chung} J.M. Chung, Phys. Lett. B {\bf 461}, 138 (1999).

\bb{AA} A. A. Andrianov, P. Giacconi and R. Soldati, JHEP {\bf0202}, 030 (2002).

\bb{cosm1} C. Adam and F.R. Klinkhamer, Nucl. Phys. B {\bf 657}, 214 (2003).

\bb{n1} F. A. Brito, L. S. Grigorio, M. S. Guimaraes, E. Passos and C. Wotzasek, Phys. Rev. D {\bf 78}, 125023 (2008).

\bb{n2} F. A. Brito, L. S. Grigorio, M. S. Guimaraes, E. Passos and C. Wotzasek, Phys. Lett. B {\bf 681}, 495 (2009). 

\bb{alt1} B. Altschul, Phys. Rev. D {\bf 69}, 125009 (2004).

\bb{baeta} A. P. B. Scarpelli, M. Sampaio, M. C. Nemes and
B. Hiller, Eur. Phys. J. C {\bf 56}, 571 (2008).




\end{thebibliography}
\end{document}